\documentclass[aps,prd,twocolumn,showpacs]{revtex4}

\usepackage{amsmath}
\usepackage{lipsum}
\usepackage{amssymb}
\usepackage{amsthm}
\usepackage{bbold}
\usepackage{dcolumn}
\usepackage{epsfig}
\usepackage{graphics}
\usepackage{graphicx}
\usepackage{color}
\usepackage{xspace}
\usepackage{cancel}
\usepackage{IEEEtrantools}
\usepackage[utf8]{inputenc}
\usepackage[colorlinks]{hyperref}

\usepackage{hyperref}
\definecolor{darkred}{rgb}{0.5,0,0}
\definecolor{darkgreen}{rgb}{0,0.5,0}
\definecolor{darkblue}{rgb}{0,0,0.5}
\hypersetup{ colorlinks,
linkcolor=darkblue,
filecolor=darkgreen,
urlcolor=darkred,
citecolor=darkblue }

\newcommand{\mi}{\mathrm{i}} 

\newcommand{\MeV}{\text{MeV}} 
\newcommand{\GeV}{\text{GeV}} 

\newcommand{\eqn}[1]{Eq.~(\ref{#1})}
\newcommand{\fig}[1]{Fig.~\ref{#1}}
\newcommand{\tab}[1]{Table~\ref{#1}}
\newcommand{\sect}[1]{Section~\ref{#1}}

\newcommand{\df}[1]{\hspace{-0.5em}\ensuremath{\frac{\mathrm{d}^{4}#1}{(2\pi)^{4}}}\,}
\newcommand{\dx}[1]{\hspace{-0.5em}\ensuremath{\mathrm{d}#1}\,}

\newcommand{\Tr}{\text{Tr}}

\begin{document}

\title{First radial excitations of heavy quarkonium in a contact interaction }

\author{Marco A. Bedolla$^{1,2}$, and Elena Santopinto$^{2}$}
\affiliation{
$^1$ Instituto de F\'{i}sica y Matem\'aticas, Universidad Michoacana de San
Nicol\'as Hidalgo, Edificio C-3, Ciudad Universitaria,
Morelia, Michoac\'an 58040, M\'exico}
\affiliation{
$^2$ Istituto Nazionale di Fisica Nucleare (INFN), Sezione di Genova, via Dodecaneso 33, 16146 Genova, Italia}

\date{\today}

\begin{abstract}

  For the flavor-singlet heavy quark systems of quarkonia, we compute the masses of the first radial excitation of mesons in four different channels: pseudo-scalar ($\eta_{c,b}(2S)$), vector ($\psi (2S),\Upsilon(2S)$),
  scalar ($\chi_{c_0,b_0}(2P)$) and axial vector ($\chi_{c_1,b_{1}}(2P)$), as well as the weak decay constants
  of the $\eta_{c,b}(2S)$ and $\psi (2S),\Upsilon(2S)$. The framework for this analysis is provided by a
  symmetry-preserving Schwinger-Dyson equations treatment of a vector$\times$vector
  contact interaction. The results found for the meson masses are in good agreement experimental data and earlier model calculations based upon Schwinger-Dyson and Bethe-Salpeter equations (BSEs) involving
  sophisticated interaction kernels. 
\end{abstract}


\keywords{Heavy quarkonium, mass spectrum, radial excitations,
  Bethe-Salpeter equation, confinement, dynamical chiral symmetry breaking, Schwinger-Dyson
  equations, contact interaction}

\maketitle

\date{\today}

\section{\label{sec:intro}Introduction}

The discovery of the $\eta_c (2S)$ in 2002~\cite{Choi:2002na,Aubert:2003pt,Asner:2003wv} was the cornerstone of a new era in meson spectroscopy. Recent advances in the understanding of quantum chromodynamics (QCD), and the recent flurry of experimental activity have led to the discovery of a bunch of new heavy states, conceiving a golden era for heavy quarkonium physics~\cite{Patrignani:2016xqp}. 

For a long time, quarkonia spectra were mainly understood from a phenomenological perspective through the use of potential models to describe the dynamics expected from QCD~\cite{Appelquist:1974zd,Appelquist:1974yr,Eichten:1974af,Appelquist:1975ya}. The potential-model approach was alluring owing to its simplicity, but became steadily more complicated and inaccurate for the heavier quarkonia states and needed new parameters fitted to data when they confronted relativist effect~\cite{Godfrey:1985xj}. However, an innovative unquenched quark model showed that relativistic effects in the coupling for bottomonia was less than $1\%$~\cite{Ferretti:2012zz,Ferretti:2013vua}, while for charmonia they were around $2-6\%$~\cite{Ferretti:2013faa}. On the other hand, direct calculations performed with Lattice QCD  successfully describe much of the quarkonia spectra~\cite{Davies:1995db,Daldrop:2011aa,Dowdall:2011wh}, though they are objectionably complex and computationally expensive.

Remarkably, in recent years, Schwinger-Dyson equations(SDEs) of quantum chromodynamics(QCD) has contributed to our understanding of these systems; their derivation makes no assumption about the strength of the interaction involved~\cite{Roberts:1994dr}. Thus, since heavy quarkonia systems connect the hard scale of the heavy constituent quarks and the soft scale of the relative momenta between them, they can conveniently to be studied through SDEs~\cite{Bashir:2012fs}.  

One of the first studies of heavy quarkonia and their radial excitation conducted through SDEs is seen in Ref.~\cite{Jain:1993qh}. Latter works with refined truncations and increased numerical complexity can be read in Refs.~\cite{Krassnigg:2004if,Bhagwat:2004hn,Bhagwat:2006xi,Maris:2005tt,Souchlas:2010zz,Blank:2011ha,Rojas:2014aka,Fischer:2014cfa,Ding:2015rkn,Raya:2016yuj,Mojica:2017tvh}. Predictions for states with exotic quantum numbers were
made in Refs.~\cite{Maris:2006ea,Krassnigg:2009zh,Qin:2011xq}, while excited mesons were investigated for quarkonia in Refs.~\cite{Jain:1993qh,Fischer:2014xha,Fischer:2014cfa,Hilger:2015hka,Hilger:2015ora,Hilger:2017jti}, and with lattice-regularized QCD~\cite{Dudek:2010wm,Liu:2012ze,Cheung:2016bym}.

The evolution of the SDEs project to the even more complicated exotic and baryonic states, decay rates and form factors is substantially non-trivial: for instance, in the calculation of elastic form factors (EFFs)~\cite{Maris:2000sk}, and transition form factors (TFF)~\cite{Chen:2016bpj}. It has been demonstrated that brute force numerical evaluation is unable to inspect the large momentum transfer region of form factors and is not considered adequate to make full comparison with already available experimental data. However, a subtle parameterization of the Bethe-Salpeter amplitudes (BSAs) in terms of Nakanishi-like perturbation theory integral representations~\cite{Nakanishi:1963zz} allows us to reach large space-like momentum transfer region~\cite{Raya:2016yuj,Chang:2013nia,Raya:2015gva}.

Recently, a symmetry-preserving vector-vector contact interaction has appeared as an alternative to full QCD-based explorations~\cite{GutierrezGuerrero:2010md,Roberts:2010rn,Roberts:2011cf,Roberts:2011wy,Chen:2012qr}. Along with the rainbow-ladder(RL) approximation of the SDEs, which is the leading order in a systematic DSE, one obtains a fully consistent treatment of the simple to implement CI model, that is helpful in providing useful results which can be compared and contrasted with full QCD calculation and experimental data. In this model, confinement is implemented by employing the proper-time regularization scheme. This scheme systematically removes quadratic and logarithmic divergences, ensuring that the axial-vector Ward-Takahashi identity (axWTI) is satisfied.

This interaction provides a good description of the masses of meson and baryon ground and
excited states for light quarks \cite{GutierrezGuerrero:2010md,Roberts:2010rn,Roberts:2011cf,Roberts:2011wy,Chen:2012qr}. The results derived from the CI model are quantitatively comparable to those obtained using sophisticated QCD model interactions,~\cite{Maris:2006ea,Cloet:2007pi,Eichmann:2008ae,Bashir:2012fs}. Strikingly fascinating, this simple CI model produces a parity-partner for each ground-state that is always more massive than its first radial excitation, so that, in the nucleon channel, e.g., the first $J^P = 1/2^-$
state lies above the second $J^P = 1/2^+$ state \cite{Chen:2012qr,Lu:2017cln}.

Building on these efforts, we extended this interaction to the analysis of the flavor-singlet heavy quarkonia systems, computing the masses of the ground-state mesons in four different channels: pseudo-scalar
($\eta_{c,b}(1S)$), vector ($J/\Psi (1S), \Upsilon(1S)$), scalar ($\chi_{c0, b0}(1P)$) and axial vector ($\chi_{c1,b1}(1P)$), as well as the weak decay constants of the $\eta_{c,b}(1S)$ and $J/\Psi (1S), \Upsilon(1S)$~\cite{Bedolla:2015mpa,Raya:2017ggu}. We also computed the EFFs of
$\eta_{c,b}(1S)$, $J/\Psi (1S)$ and $\Upsilon (1S)$. Additionally, we calculated the transition form factor for $\eta_{c,b} \to\gamma^*\gamma$~\cite{Bedolla:2016yxq,Raya:2017ggu}. In these previous works, we found that the contact interaction provides form factors that are harder than those expected from a proper treatment of full QCD with a running mass function.

Our intention is to implement that model~\cite{Bedolla:2015mpa,Raya:2017ggu} on the calculation of spectra of radially excited heavy-quarkonia states. We compute their mass spectrum by using the dimensionless coupling for the CI model present in Ref.~\cite{Raya:2017ggu}.  In addition, we compute their weak decay constants, whose precise knowledge is of huge importance for the hadronic observables measured by LHCb and FAIR-GSI, for example.

This paper is organized as follows: in~\sect{sec:ci-mod} we give the minimum details necessary to
the SDE-BSE approach to mesons, employing the contact interaction in the RL approximation, and the consequences that this interaction has for the interaction kernels. In Section~\ref{sec:radialexcitations}, the model is minimally modified to calculate the first radial excitation. 
In~\sect{sec:massspectrum}, we tabulate our results for the mass spectrum of ground state
quarkonia and the decay constants for $\eta_{c,b}(2S)$, $\Psi (2S)$ and $\Upsilon(2S)$ in the minimally modified
CI model. Finally, in \sect{sec:conclusions}, we present our conclusions.

\section{\label{sec:ci-mod} Contact interaction model}

We dedicate this section to recapitulating the CI model and its implementation in a SDE-BSE formalism to study two-particle bound systems. We precise their connection with chiral symmetry breaking and confinement; and  we describe the method adopted to obtain the results disclosed in this article. For a more detailed description of the model, see Refs.~\cite{Chen:2012qr,Bedolla:2015mpa}.

\subsection{SDE-BSE formalism}

The $f$-flavor dressed-quark propagator $S_{f}$ is obtained by solving the quark
SDE~\cite{Alkofer:2000wg,Maris:2003vk,Holl:2006ni,Roberts:2007jh}
\begin{eqnarray}
 \label{eqn:quark_sde}
 &&\hspace{-0.5cm} S_{f}^{-1}(p)=i\gamma\cdot p + m_{f} + \Sigma_{f}(p), \\
 \label{eqn:quark_se}
 && \hspace{-0.5cm} \Sigma_{f}(p)=
\int\df{q}
g^{2}D_{\mu\nu}(p-q)\frac{\lambda^{a}}{2}\gamma_{\mu}S_{f}(q)\Gamma^{a}_{\nu}(p,q),
\end{eqnarray}
\noindent where $g$ is the strong coupling constant, $D_{\mu\nu}$ is the dressed-gluon propagator,
$\Gamma^{a}_{\nu}$ is the dressed-quark-gluon vertex, and $m_{f}$  is the $f$-flavor current-quark
mass.

Since the SDEs form a coupled infinite set of non linear integral equations, a truncation scheme
is required in order to characterize a tractable problem. This is achieved once we have specified the gluon propagator and the quark-gluon vertex. For a comprehensive recent review of the SDE-BSE
formalism and its applications to hadron physics, see, for example, Ref.~\cite{Bashir:2012fs}.

Because high-energy experiments cannot perceive quarks directly, bound states have to be studied in order to test QCD. Specifically, a meson bound-state problem in an explicit $J^{PC}$ channel is determined by its homogeneous BSE~\cite{Gross:1993zj,Salpeter:1951sz,GellMann:1951rw},
\begin{equation}
\label{eqn:bse}
\left[\Gamma_{H}(p;P)\right]_{tu}=
\int\df{q}K_{tu;rs}(p,q;P)\chi(q;P)_{sr},
\end{equation}
\noindent where $\chi(q;P)=S_{f}(q_{+})\Gamma_{H}(q;P)S_{g}(q_{-})$ is the Bethe-Salpeter wave-function; $q_{+}=q+\eta P$,
$q_{-}=q-(1-\eta)P$; $\eta \in [0,1]$ is a momentum-sharing parameter, $p$ ($P$) is the relative
(total) momentum of the quark-antiquark system; $S_{f}$ is the $f$-flavor dressed-quark propagator; 
$\Gamma_{H}(p;P)$ is the meson Bethe-Salpeter amplitude (BSA), where $H$ specifies the quantum
numbers and flavor content of the meson; $r,s,t$, and $u$ represent color, Dirac and flavor indices; and $K(p,q;P)$ is the quark-antiquark scattering kernel. 
They also specify the kernel in the BSE,
\eqn{eqn:bse}, through the axial-vector Ward-Takahashi identity (axWTI)~\cite{Maris:1997hd}
\begin{equation}
  \label{eqn:axwti} 
  -\mi P_{\mu}\Gamma_{5\mu}(k;P)=S^{-1}(k_{+})\gamma_{5} + \gamma_{5}S^{-1}(k_{-}).
\end{equation}
\noindent Equation~(\ref{eqn:axwti}), which encodes the phenomenological features of dynamical
chiral symmetry breaking (DCSM) in QCD, relates the axial-vector vertex, $\Gamma_{5\mu}(k;P)$, to the quark 
propagator, $S(k)$. This in turn implies a relationship between the kernel in the BSE,
\eqn{eqn:bse}, and that in the quark SDE, \eqn{eqn:quark_sde}. 
This relation must be preserved by any viable truncation scheme of the SDE-BSE coupled system,
thus constraining the content of the quark-antiquark scattering kernel $K(p,q;P)$.

\subsection{Rainbow-ladder truncation and the contact interaction}
In a symmetry-preserving vector$\times$vector contact interaction, one considers that the interaction between quarks is not mediated via massless bottom exchange, but instead through the interaction defined by
\begin{eqnarray}
\label{eqn:contact_interaction}
g^{2}D_{\mu \nu}(k)&=&\frac{4\pi\alpha_{\text{IR}}}{m_g^2}\delta_{\mu \nu} \equiv
\frac{1}{m_{G}^{2}}\delta_{\mu\nu}, \\
\label{eqn:quark_gluon_vertex_rl}
\Gamma^{a}_{\mu}(p,q)&=&\frac{\lambda^{a}}{2}\gamma_{\mu},
\end{eqnarray}
\noindent where $m_g=800\,\MeV$ is a gluon mass scale which is in fact generated dynamically in
QCD~\cite{Boucaud:2011ug}, and $\alpha_{\text{IR}}$  is the CI model parameter, which can be interpreted as the interaction strength in the infrared~\cite{Binosi:2016nme,Deur:2016tte}. The simultaneous implementation of \eqn{eqn:quark_gluon_vertex_rl} for the quark-gluon vertex and \eqn{eqn:bskernel_rl_contact}
for the scattering kernel is the familiar rainbow-ladder approximation.

Once the kernel has been specifed by Equations~(\ref{eqn:contact_interaction}) and (\ref{eqn:quark_gluon_vertex_rl}) in the quark SDE, \eqn{eqn:quark_sde}, then the  the general form of the momentum-independent $f$-flavored dressed-quark propagator within the context of the rainbow-ladder truncation and a contact interaction is~\cite{GutierrezGuerrero:2010md, Roberts:2010rn,
Roberts:2011cf,Roberts:2011wy,Chen:2012qr,Bedolla:2015mpa,Bedolla:2016yxq}
\begin{equation}
\label{eqn:quark_inverse_contact}
S_{f}^{-1}(p)= \mi\gamma\cdot p + M_{f}\,.
\end{equation}
Thus, the flavor-dependent fermion constant mass $M_{f}$ is obtained by solving
\begin{equation}
  \label{eqn:const_mass} M_{f} = m_{f} +
  \frac{16M_{f}}{3\pi^{2}m_{G}^{2}}\int\df{q}\frac{1}{q^{2}+M_{f}^{2}}.
\end{equation}

Since the integral in \eqn{eqn:const_mass} is divergent, we must adopt a regularization procedure.
We employ the proper time regularization scheme~\cite{Ebert:1996vx} to write \eqn{eqn:const_mass}
as
\begin{eqnarray}
  \label{eqn:const_mass_reg} 
 &&M_{f}= m_{f} + \frac{M_{f}^3}{3\pi^{2}m_{G}^{2}}
  \Gamma(-1,\tau_{\text{UV}} M_{f}^2,\tau_{\text{IR}} M_{f}^2)\,,
  \label{eqn:Ifun}
\end{eqnarray}
\noindent where $\Gamma(a,z_{1},z_{2})$ is the generalized incomplete Gamma function:
\begin{equation}
\label{eqn:incomplete_gamma}
\Gamma (a, z_1,z_2)=\Gamma (a,z_1)-\Gamma(a,z_2)\,.
\end{equation}
The parameters $\tau_{\text{IR}}$ and $\tau_{\text{UV}}$ are infrared and ultraviolet
regulators, respectively. A nonzero value for  $\tau_{\text{IR}}\equiv 1/\Lambda_{\text{IR}}$ implements 
confinement~\cite{Roberts:2007ji}. Since the CI is nonrenormalizable theory, 
$\tau_{\text{UV}}\equiv 1/\Lambda_{\text{UV}}$ becomes part of the model and therefore sets the scale for
all dimensional quantities. 
The importance of an ultraviolet cutoff in Nambu--Jona-Lasinio-type models has also been discussed
in Refs.~\cite{Farias:2005cr,Farias:2006cs}.

In the context of the contact interaction and rainbow-ladder truncation, \eqn{eqn:bse} gives
\begin{equation}
  \label{eqn:bskernel_rl_contact} K(p,q;P)= -g^{2}D_{\mu\nu}(p-q)
  \left[\frac{\lambda^{a}}{2}\gamma_{\mu}\right]\otimes
  \left[\frac{\lambda^{a}}{2}\gamma_{\nu}\right],
\end{equation}
\noindent where $g^{2}D_{\mu\nu}$ is given by \eqn{eqn:contact_interaction}.
Thus, the homogeneous BSE ($\eta=1$) takes the simple form
\begin{equation}
  \label{eqn:bse_contact} 
  \Gamma_{H}(p;P)=-\frac{4}{3}\frac{1}{m_{G}^{2}}\int\df{q}
  \gamma_{\mu}S_{f}(q+P)\Gamma_{H}(q;P)S_{g}(q)\gamma_{\mu}.
\end{equation}
Furthermore, the implementation of axWTI entails~\cite{GutierrezGuerrero:2010md,Roberts:2010rn,Roberts:2011cf,
  Roberts:2011wy,Chen:2012qr,Bedolla:2015mpa,Bedolla:2016yxq}
\begin{equation}
  \label{eqn:wticorollary}
 0=\int_{0}^{1}\dx{x}\int\df{q} \frac{\frac{1}{2}q^{2} +
    \mathfrak{M}^{2}} {\left(q^{2}+ \mathfrak{M}^{2}\right)^{2}},
\end{equation}
where $$\mathfrak{M}^{2}=M_{f}^{2}x + M_{g}^{2}(1-x)+ x(1-x)P^{2}\,.$$ 	
Equation~(\ref{eqn:wticorollary}) states that the axWTI is satisfied if, and only if, the model
is regularized so as to ensure there are no quadratic or logarithmic divergences: circumstances
under which a shift in integration variables is permitted, an operation required in order to prove 
\eqn{eqn:axwti}~\cite{GutierrezGuerrero:2010md,Roberts:2010rn,Roberts:2011cf,Roberts:2011wy,
  Chen:2012qr}.

Since the interaction kernel given by \eqn{eqn:bskernel_rl_contact} does not depend on the external
relative momentum, a symmetry-preserving regularization will give momentum independent solutions. In this case, the general forms of the BSAs for the pseudoscalar, scalar, vector, and 
axial-vector channels, respectively, are given by
\begin{eqnarray}
  \label{eqn:psbsagral}
  \Gamma^{0^{-}}(P)&=& \gamma_{5}\left[\mi E^{0^{-}}(P)
    + \frac{1}{2M}\gamma\cdot P F^{0^{-}}(P)\right], \\
  \label{eqn:sbsagral}
  \Gamma^{0^{+}}(P)&=& \mathbb{1}E^{0^{+}}(P), \\
  \label{eqn:vbsagral}
  \Gamma^{1^{-}}_{\mu}(P)&=&\gamma^{T}_{\mu}E^{1^{-}}(P)
  + \frac{1}{2M}\sigma_{\mu\nu}P_{\nu}F^{1^{-}}(P), \\
  \label{eqn:avbsagral}
  \Gamma^{1^{+}}_{\mu}(P)&=&\gamma_{5}\left[\gamma^{T}_{\mu}E^{1^{+}}(P)
    + \frac{1}{2M}\sigma_{\mu\nu}P_{\nu}F^{1^{+}}(P)\right],
\end{eqnarray}
\noindent where $M$ is a mass scale, built from solutions of \eqn{eqn:const_mass_reg}. Results for
physical observables are independent of this choice.

Since the BSE is a homogeneous equation, the BSA has to be normalized by a separate equation.
In the rainbow-ladder approximation, this condition is
\begin{equation}
  \label{eqn:RLNorm}
  1=N_{c}\frac{\partial}{\partial P^{2}}\int\df{q}
  \Tr\left[\overline{\Gamma}_{H}(-Q)S(q+P)\Gamma_{H}(Q)S(q)\right],
\end{equation}
\noindent evaluated at $Q=P$, where $P^{2}=-m_{H}^{2}$, $\Gamma_{H}$ is the normalized BSA, and
$\overline{\Gamma}_{H}$ is its charge-conjugated version. For the vector and axial-vector channels,
there is an additional factor of 1/3 on the right-hand side to account for all three polarizations
of a spin-1 meson.

Once the BSA has been normalized, observables can be computed. The pseudoscalar and vector
meson decay constants, $f_{0^{-}}$ and $f_{1^{-}}$, are defined, respectively, by
\begin{eqnarray}
  \label{eqn:psdecaydef}
  &&\hspace{-2em}P_{\mu}f_{0^{-}}=
  N_{c}\int\df{q}\Tr\left[\gamma_{5}\gamma_{\mu}
    S(q_{+})\Gamma_{0^{-}}(P)S(q_{-})\right], \\
  \label{eqn:vdecaydef}
  &&\hspace{-2.75em}m_{1^{-}}f_{1^{-}}=
  \frac{N_{c}}{3}\int\df{q}\Tr\left[\gamma_{\mu}
    S(q_{+})\Gamma^{1^{-}}_{\mu}S(q_{-})\right],
\end{eqnarray}
\noindent where $m_{1^{-}}$ is the mass of the vector bound state, and the factor of 3 in the
denominator on the right-hand side of \eqn{eqn:vdecaydef} comes from summing over the three
polarizations of the spin-1 meson.

\subsection{\label{sec:unifiedCImodel} CI running coupling}

%
\begin{table}[h]
\begin{center}
\begin{tabular}{lllll}
\hline \hline
 quark & $\hat{\alpha}_{\mathrm IR}\;[\GeV^{-2}]$ & $\Lambda_{\mathrm UV}\;[\GeV] $ & $\alpha$ & Ratio \\
\hline
$u,d,s$ & 4.565 & 0.905 & 3.739 & 1 \\
$c$     & 0.228 & 2.400 & 1.547 & 0.414 \\
$b$     & 0.035 & 6.400 & 1.496 & 0.400 \\
\hline \hline
\end{tabular}
\caption{\label{tab:rc} Dimensionless coupling constant $\alpha=\hat{\alpha}_{\mathrm IR}\Lambda_{\mathrm UV}^{2}$,
  where $\hat{\alpha}_{\mathrm IR}=\alpha_{\mathrm IR}/m_{g}^{2}$, for the contact interaction, extracted from a
  best-fit to data, as explained in Ref.~\cite{Raya:2017ggu}. Fixed parameters are $m_{g}=0.8\,\GeV$ and
  $\Lambda_{\mathrm IR}=0.24\,\GeV$.}
\end{center}
\end{table}
%

%
In a previous paper~\cite{Raya:2017ggu}, we explained how the CI can be used to study light and heavy mesons. When studying the  heavy sector, a change in the model parameters has to be made: an increase in the ultraviolet regulator, and a reduction in the coupling strength. Subsequently, we figured out that different set parameters are needed in order to study each sector: light, charm and bottom, as displayed in~\tab{tab:rc}. With these parameters, we define a dimensionless coupling $\alpha$ guided by~\cite{Farias:2005cr,Farias:2006cs}
\begin{equation}
\alpha=\frac{\alpha_{IR}}{m_{g}^{2}}\Lambda_{\mathrm UV}^2.
\end{equation}
\noindent The drop in $\alpha$, in relation to its value in the light-quarks sector, can be
read off from the last column of \tab{tab:rc}. Indeed, $\alpha$ is reduced by a factor of
$2.1-2.3$ on going from the light to the heavy sector, instead of the apparent large factors
quoted in Tables~\ref{tab:mcc_all_opt} and~\ref{tab:mbb_all_opt}.

Moreover, as a reminiscent of the running coupling QCD with the momentum scale at which it is measured, an inverse logarithmic curve can fit reasonably well the functional dependence of $\alpha(\Lambda_{\mathrm UV})$. The fit reads
\begin{equation}
\label{eqn:logaritmicfit} \alpha(\Lambda_{\mathrm UV})=a\ln^{-1}\left(\Lambda_{\mathrm UV}/\Lambda_0\right) \,,
\end{equation}
where $a=0.923$ and $\Lambda_0=0.357$ ~\cite{Raya:2017ggu}. With this fit, it is viable to recover the value of the strength coupling $\alpha$ once given a value of $\Lambda_{\mathrm UV}$. We will apply this feature in the future.
%
\begin{figure}[ht]
\includegraphics[width=0.45\textwidth]{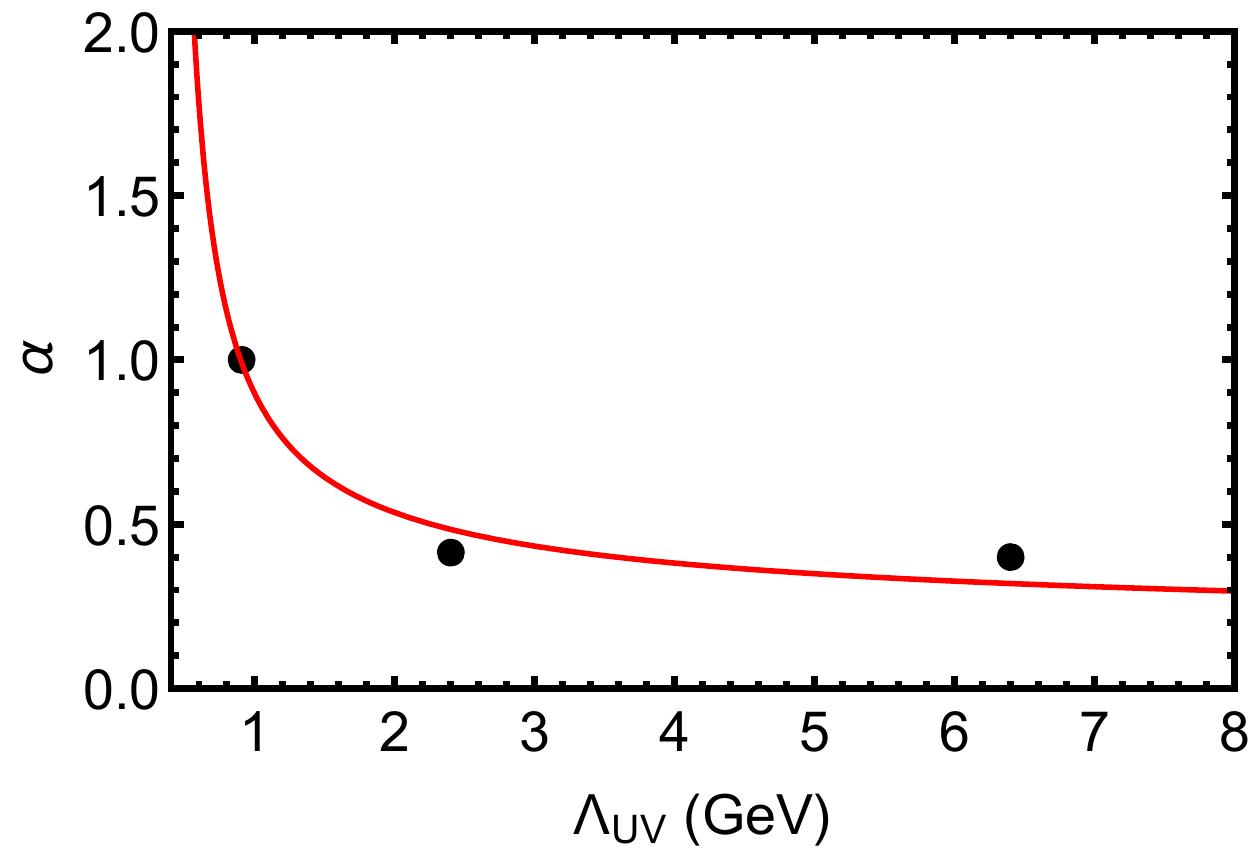}
\caption{\label{fig:dimensionlesscoupling} Dimensionless coupling $\alpha$ for the contact interaction. It is interesting to note that the variation of the coupling $\alpha$ as a function of $\Lambda_{\mathrm UV}$ is not far from a logarithmic decrease fitted by the equation.}
\end{figure}
%

\section{\label{sec:radialexcitations} Meson first radial excitation in a contact interaction}
\begin{figure}[ht]
\includegraphics[width=0.45\textwidth]{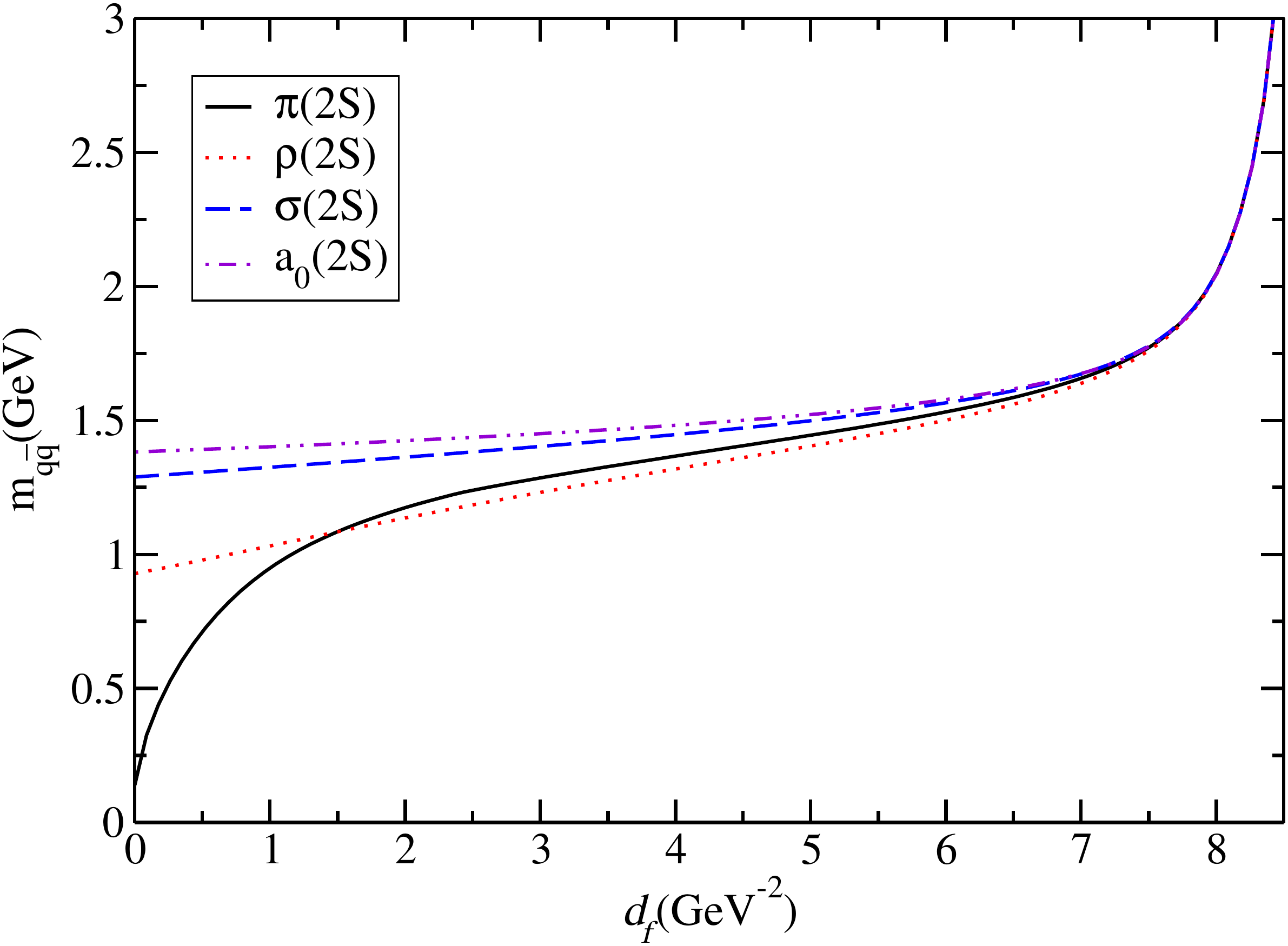}
\caption{\label{fig:pionexcited} Contact interaction evolution for $u\bar u$ first radially excited states in terms of $d_{\cal F}$. The upper bound is found at $d_{\cal F}=8.7\,\mathrm{GeV}^{-2}$, where the value of the masses diverges. We use $u, d, s$ parameters from \tab{tab:rc}. }
\end{figure}

Recently, we have developed a practical CI model suited to calculate observables for light and heavy mesons~\cite{Raya:2017ggu}. Another paper has focused on $D$ mesons using this model~\cite{Serna:2017nlr}. These studies have shown that as the system becomes heavier, a decrease in the coupling and an increase in the ultraviolet cut-off are required. Additionaly, observables for first radial excitations using this model have been already calculated in the light sector~\cite{Roberts:2011wy,Chen:2012qr}. As a continuation of these works, we will adopt their strategy as a starting point from which to calculate first radial excitations in heavy quarkonia systems.

Generally, most of the studies that use the SDE-BSE formalism analyze only ground-state systems. When examining radially excited states in fully covariant approaches, a single zero is related to the first radially excited bound-state. This zero is usually seen in the relative-momentum dependence of the leading Tchebychev moment of its dominant Dirac structure~\cite{Qin:2011xq,Holl:2004fr,Holl:2005vu, Segovia:2015hra}. However, this approximation is not exempt from problems: as the analytic structure of the quark propagator restricts the range of direct calculation~\cite{Bhagwat:2002tx}, some extrapolation may be needed~\cite{Bhagwat:2007rj}; additionally, some non-physical solutions may show up, and these cannot be easily related to a physical state~\cite{Ahlig:1998qf}.

Furthermore, since a single zero cannot be exhibited by a momentum-independent relative-momentum bound-state amplitude, the possibility that the interaction between quarks is momentum-independent vanishes. In other words: if the zero is established at $k_0^2$, then a momentum-independent interaction can only produce reliable results for phenomena that probe momentum scales $k^2\ll k_0^2$.  In the light sector $k_0^2 \sim 2 M^2 \sim (0.5\,{\rm GeV})^2$ is mostly found~\cite{Qin:2011xq,Holl:2004fr,Holl:2005vu}. However, in Refs.\,\cite{Volkov:1996br,Volkov:1999xf}, a way out of this predicament is indicated: in the phenomenological analysis of the contact interaction, a zero by hand is inserted into the analytical expressions of the kernels built up from  Eqs.~\eqref{eqn:bse}.  Now we identify the BSE for a radial excitation as the form of Eq.\,(\ref{eqn:bse_contact}) obtained with Eq.\,(\ref{eqn:contact_interaction}) and insert a node by hand into the BSE; this reads \cite{Roberts:2011cf,Chen:2012qr}
\begin{equation}
\label{eqn:locatezero}
\Gamma_{H_1}=\Gamma_{H}\left(1 - d_{\cal F} q^2\right)\,,
\end{equation}
which forces a zero into the kernel at $q^2=1/d_{\cal F}$, where $d_{\cal F}$ is a parameter that will be specified later.  Essentially, the mass of the bound-state is increased because the presence of this zero reduces the coupling in the BSE. The mass of a meson radial excitation is elevated by an equivalent mechanism when employing more sophisticated interactions kernels. The presence of this zero needs te replacement $\mathcal{C}_{\alpha\beta}\to \mathcal{F}_{\alpha\beta}$ of each BSE kernel found in Appendix A of Ref.~\cite{Bedolla:2015mpa}
\begin{IEEEeqnarray}{rCl}
{\cal F}_{\alpha\beta}(\mathfrak{M}^{2})
&=& {\cal C}_{\alpha\beta}(\mathfrak{M}^{2})
- d_{\cal F} {\cal D}_{\alpha\beta}(\mathfrak{M}^{2})\,,\\
{\cal D}_{01}(\mathfrak{M}^{2}) & = & \int_0^\infty ds\,s^2\,\frac{1}{s+\mathfrak{M}^{2}}
\nonumber\\
&\to&  \int_0^\infty ds\,s^2\int_{\tau_{UV}^2}^{\tau_{IR}^2} 
\exp\left[-\tau \left(s+\mathfrak{M}^{2}\right)\right].\label{eqn:excited_reg}
\end{IEEEeqnarray}
Because the renormalization condition \eqn{eqn:RLNorm} involves two amplitudes given by \eqn{eqn:locatezero}, for an excited state we substitute $\mathcal{C}_{\alpha\beta}\to \mathcal{G}_{\alpha\beta}$ in each renormalization condition in Appendix B of Ref.~\cite{Bedolla:2015mpa}
\begin{IEEEeqnarray}{rCl}
{\cal G}_{\alpha\beta}(\mathfrak{M}^{2})
&=& {\cal C}_{\alpha\beta}(\mathfrak{M}^{2})
- 2 d_{\cal F} {\cal D}_{\alpha\beta}(\mathfrak{M}^{2})+ d_{\cal F} ^2 {\cal E}_{\alpha\beta}(\mathfrak{M}^{2})\nonumber\,,\\\\
{\cal E}_{01}(\mathfrak{M}^{2}) & = & \int_0^\infty ds\,s^3\,\frac{1}{s+\mathfrak{M}^{2}}
\nonumber\\
&\to&  \int_0^\infty ds\,s^3\int_{\tau_{UV}^2}^{\tau_{IR}^2} 
\exp\left[-\tau \left(s+\mathfrak{M}^{2}\right)\right],\label{eqn:excited_reg2}
\end{IEEEeqnarray}
where ${\cal D}_{01}$, ${\cal E}_{01}$ and related terms are explicitly given in Appendix~\ref{App:ExcitedRegularization}.

\begin{figure}[ht]
\includegraphics[width=0.45\textwidth]{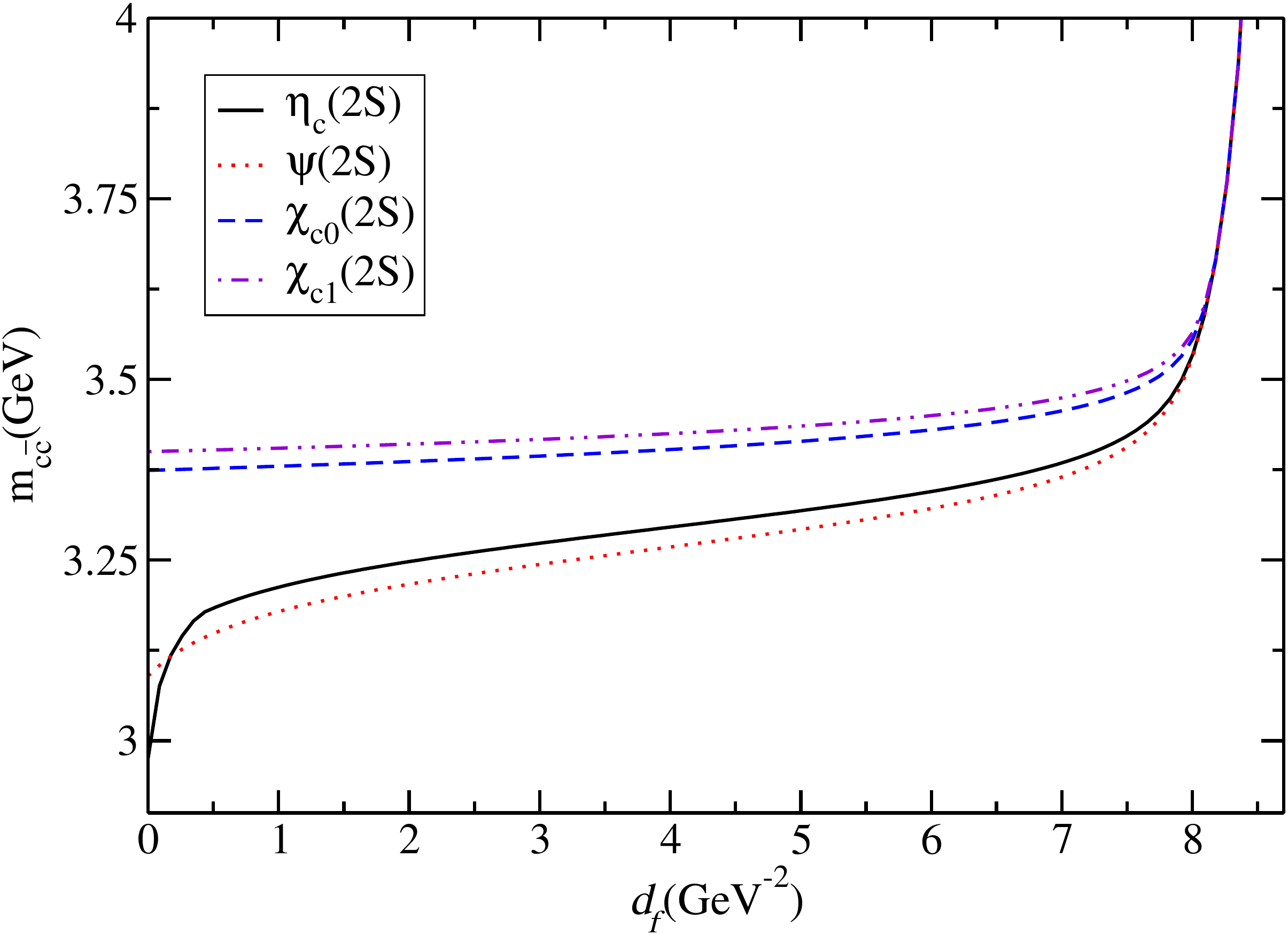}
\caption{\label{fig:charmexcited} Contact interaction evolution for $c\bar c$ first radially excited states in terms of $d_{\cal F}$. The upper bound is found at $d_{\cal F}=8.7\,\mathrm{GeV}^{-2}$, where the value of the masses diverges. We use $c$ parameters from \tab{tab:rc}.}
\end{figure}

\begin{figure}[ht]
\includegraphics[width=0.45\textwidth]{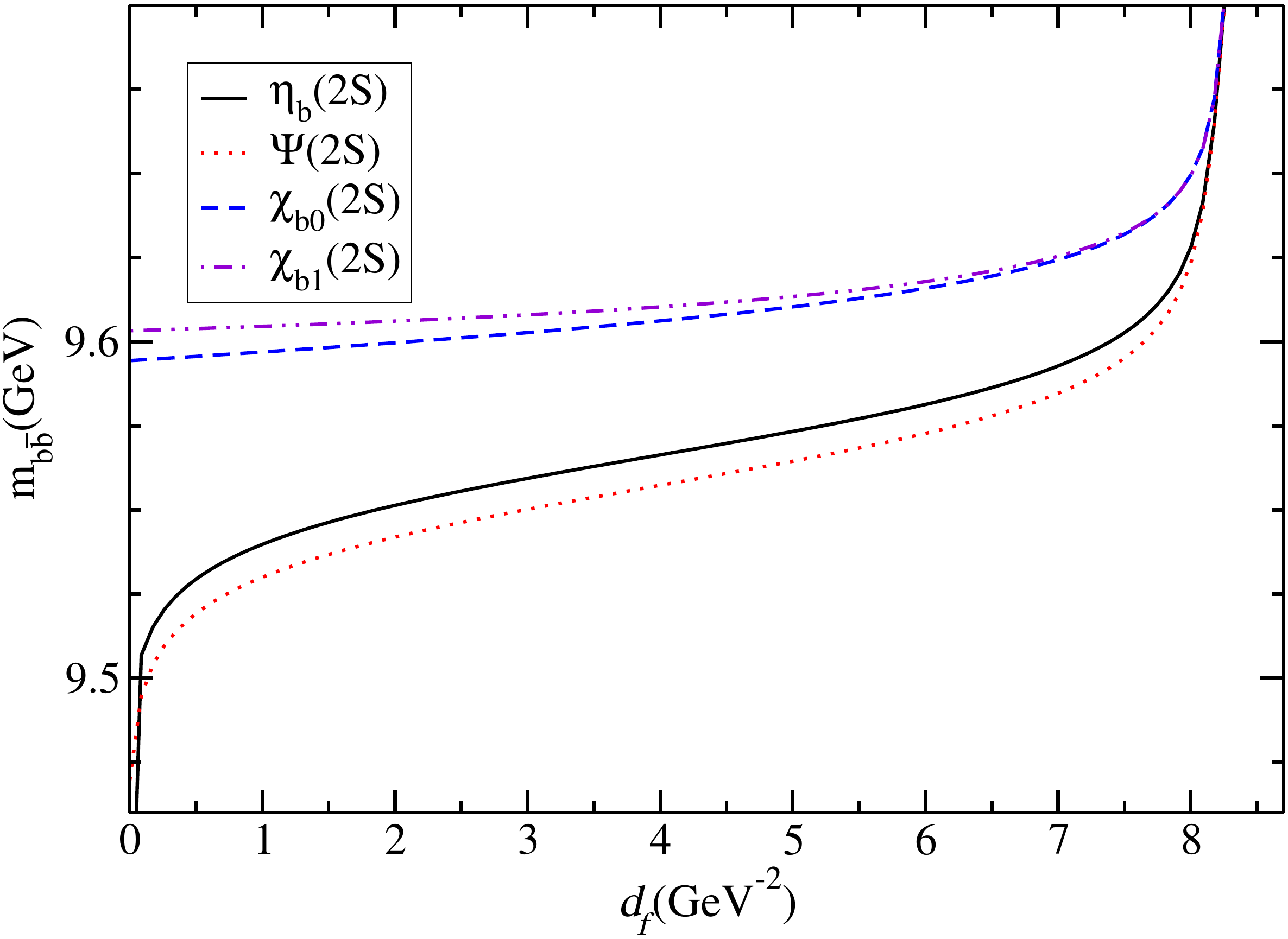}
\caption{\label{fig:bottomexcited} Contact interaction evolution for $b\bar b$ first radially excited states in terms of $d_{\cal F}$. The upper bound is found at $d_{\cal F}=8.7\,\mathrm{GeV}^{-2}$, where the value of the masses diverges. We use $b$ parameters from \tab{tab:rc}.}
\end{figure}

Now, it only remains to specify the choice of the new parameter $d_{\cal F}$ introduced in \eqn{eqn:locatezero}. In \fig{fig:pionexcited}, we show the evolution with $d_f$ on the excited states masses from the light sector. We immediately notice that there is a switch between the ordering of $m_{\pi^*}$ and $m_{\rho^*}$ at 1.5 GeV$^{- 2}$. Additionally, there is also a maximum value for $d_f$ at 8.7 GeV$^{- 2}$, where the value of the mass for all channels diverges. This maximum value is independent on the current quark mass and the bounds restricted by the infrared and ultraviolet cutoffs, as appears in Figs.~\ref{fig:charmexcited}-\ref{fig:bottomexcited}.

On the other hand, the zero in the BSA, namely $z_{q\bar q}^*$, must lie in the domain [$\Lambda_{\mathrm IR}^2$,$\Lambda_{\mathrm UV}^2$]. Thus, the minimal value should be $d_{\cal F}^{\mathrm{min}}=1/\Lambda_{\mathrm UV}^2=1.22,\,0.173,$ and $0.024\,\mathrm{GeV}^{-2}$ for light, charm and bottom sectors, respectively. Likewise, the maximum value should be at $d_{\cal F}^{\mathrm{max}}=1/\Lambda_{\mathrm IR}^2=17.36\,\mathrm{GeV}^{-2}$ for all regions; this value does not contradicts the maximum value found above. As can be seen from the figures, the bound-state mass increases when $d_{\cal F}$ rises, this indicates that the energy is stored in the zero crossing: the more the domain  [$\Lambda_{\mathrm IR}^2$,$z_{q\bar q}^*$] is compressed, the more massive is the excited state in relation to its ground-state. 
Following Refs.~\cite{Roberts:2011wy,Chen:2012qr}, we choose $1/d_{\cal F}=8 M_R ^2$, where $M_R$ is the reduced mass
\begin{equation}
\label{eqn:reducedmass}
M_R=\frac{M_f M_g}{M_f+M_g}\,,
\end{equation}
to obtain a value for $m_{\pi(2S)}$=1.33GeV. In the next section we test this scheme and calculate the spectrum of first radially excited heavy mesons  states.

\section{\label{sec:massspectrum} Numerical Results}

Tables~\ref{tab:muu_all_opt}-\ref{tab:mbb_all_opt} show the masses of ground-states and first radial excitations for different channels that have been studied with the CI. Results of ground-states were obtained and are to be found in the discussion in previous papers~\cite{Roberts:2011wy,Chen:2012qr,Bedolla:2015mpa,Raya:2017ggu}; and the light-sector analysis of radial excitations is found in~\cite{Roberts:2011wy,Chen:2012qr}.

Our results are displayed in Tables~\ref{tab:mcc_all_opt} and \ref{tab:mbb_all_opt}. We notice that, in maintaining the set of parameters used for the ground-states, then radially excited states of pseudoscalar channels are more massive than those of their respective vector channels, this means $m_{0^{-+}}^*>m_{1^{--}}^*$. In contrast, experimental results indicate $m_{0^{-+}}^*<m_{1^{--}}^*$. This incorrect ordering also appears in fully-covariant models~\cite{Qin:2011xq} when analysing the light sector; therefore it is considered a weakness of the rainbow-ladder approximation. Furthermore, as an indication of a naive application to the heavy sector, the first radially excited masses obtained are even lower than their ground-state counterparts in all but the pseudoscalar channel.

In the phenomenological analysis of the contact interaction, in order to obtain an experimental value for the $a_1-\rho$ splitting, a spin-orbit repulsion has been introduced into the scalar and axial-vector channels to simulate a large dressed-quark anomalous chromomagnetic moment~\cite{Roberts:2011wy,Chen:2012qr}. In a recent study, when including opposite-parity diquark correlation in the analysis of the structure of baryons, a second spin-orbit parameter is suggested to match the $a_1-\rho$ and $\sigma-\rho$ splitting produced by sophisticated Bethe-Salpeter kernels. The introduction of these parameters simulates the DCSB effects that are crucial for a successful description of the meson spectrum with truncations beyond RL~\cite{Lu:2017cln}. However, the implementation of that spin-orbit coupling into the BSE kernel has negligible effects in the heavy sector~\cite{Bedolla:2015mpa}. 

The analysis of heavy mesons by means of the CI requires a diminution of the coupling accompanied by an increase in the ultraviolet cut-off, as explained in detail in Ref.~\cite{Bedolla:2015mpa}. Precisely, when the mass of the heavy quarks increases, the further the coupling between them decreases and has to be compensated by an increment in the ultraviolet coupling. On the other hand, in fully covariant models, when the meson mass increases, the effective coupling decreases, and a small correction of the mass may mean a large correction of the binding energy~\cite{Jain:1993qh}. 

The sensitivity to the coupling in the heavy sector explains the small values that we obtained in our first attempt to calculate first radially excited heavy quarkonia states with the same parameters as their ground states. Additionally, in pursuing a fix in $d_{\cal F}$ to obtain a correct value for $\eta_c (2S)$ and $\eta_c (2S)$, we get too close at the divergence value of 8.7 GeV$^{-2}$, as it can be seen in Figs.~\ref{fig:charmexcited}-\ref{fig:bottomexcited}. This means that a small variation on $d_{\cal F}$ is reflected in a big change on the value of the mass and, as a consequence, we have to present a proposal to study first radially excited states with the CI.

\begin{figure}[ht]
\includegraphics[width=0.45\textwidth]{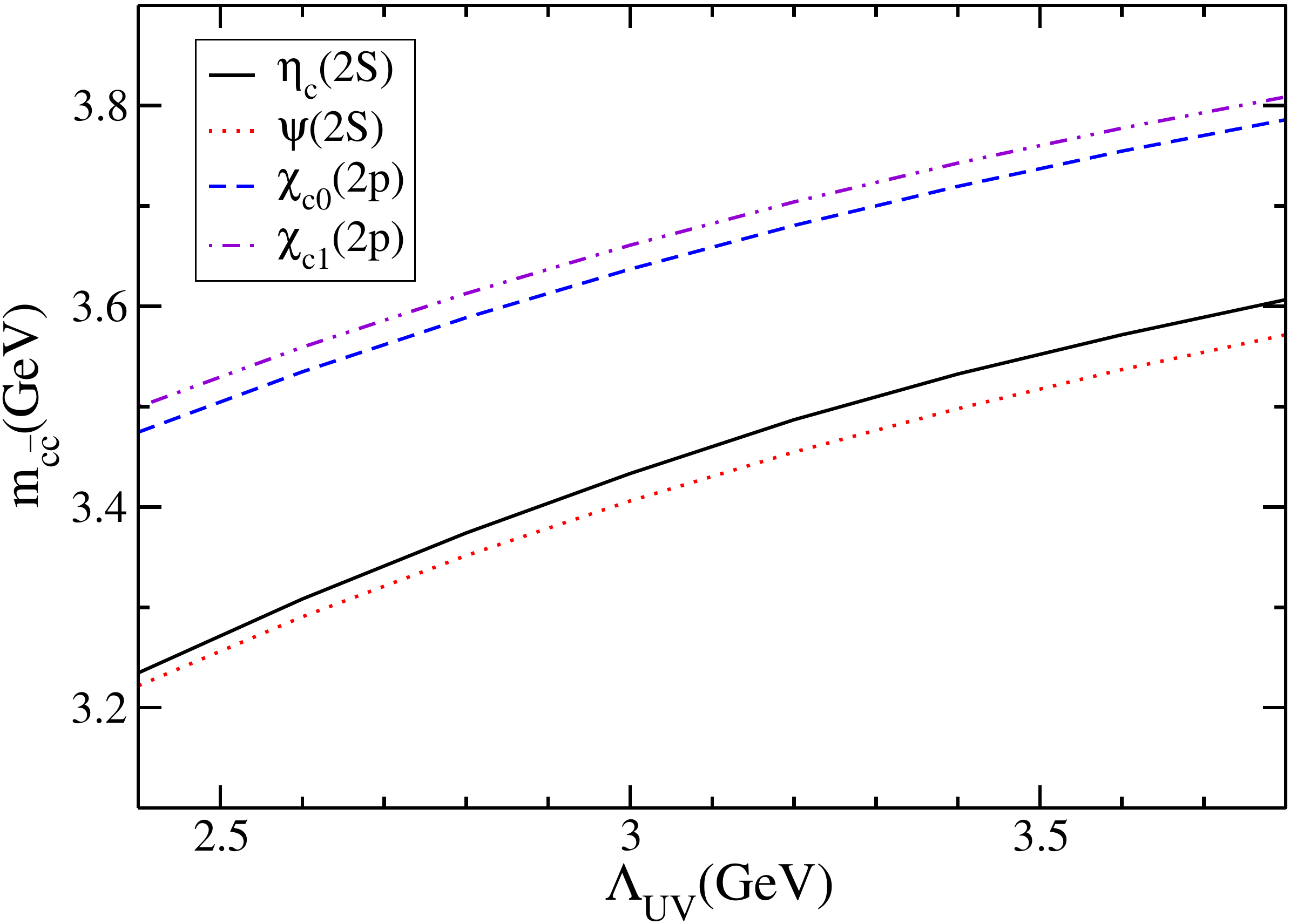}
\caption{\label{fig:charmoniumex} Contact interaction evolution for $c\bar c$  first radially excited states in terms of $\Lambda_{\mathrm UV}$.}
\end{figure}

\begin{figure}[ht]
\includegraphics[width=0.45\textwidth]{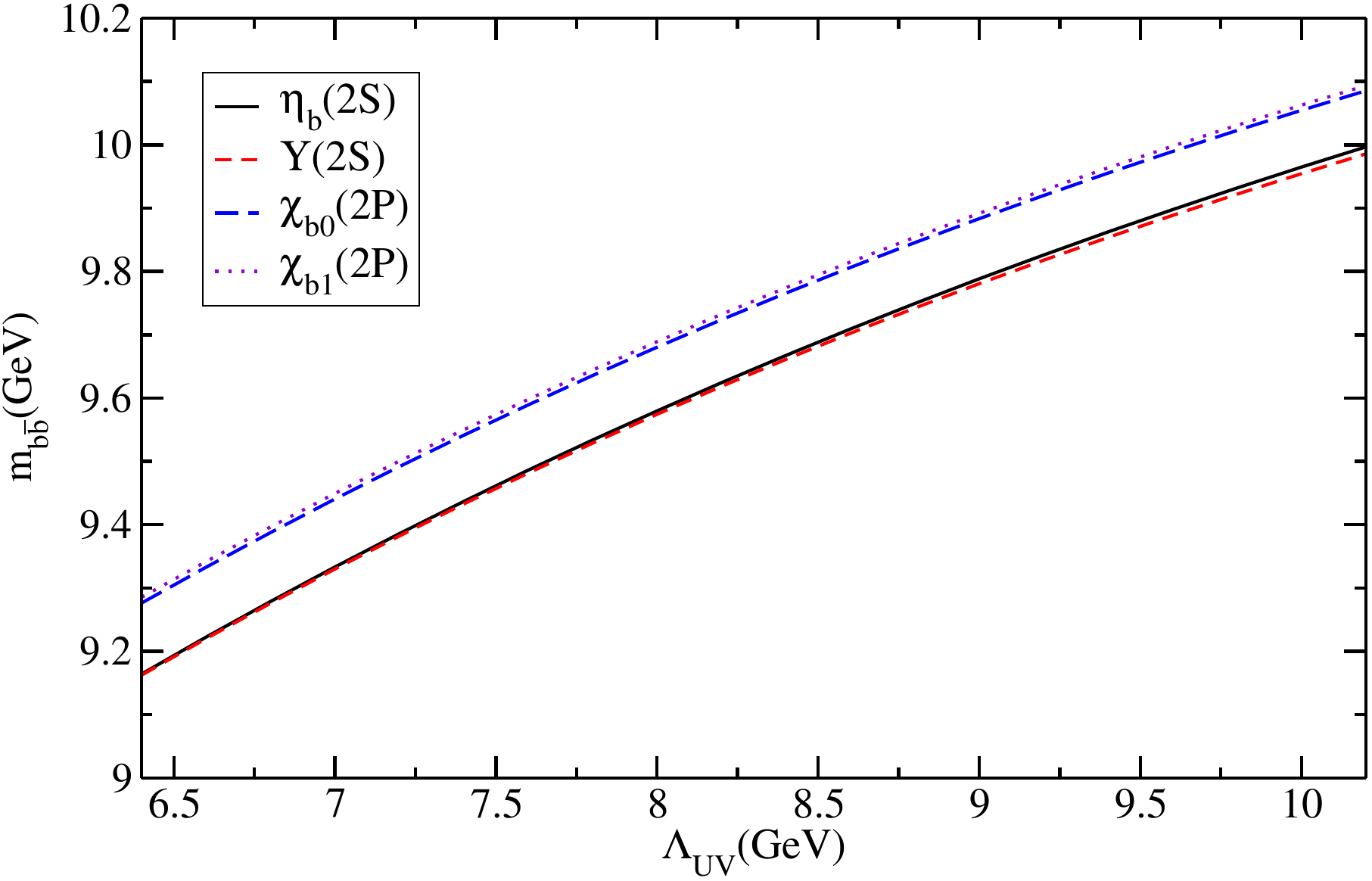}
\caption{\label{fig:bottomoniumex} Contact interaction evolution for $b\bar b$ first radially excited states in terms of $\Lambda_{\mathrm UV}$.}
\end{figure}

With the aid of \eqn{eqn:logaritmicfit}, we calculate the evolution of quarkonia first radially excited  states masses in relation to $\Lambda_{\mathrm UV}$: charmonia spectra are shown in~ \fig{fig:charmoniumex}, while those of bottomonia are displayed in \fig{fig:bottomoniumex}. They show a slight increase in the masses, as a result of the evolution of the quark-dressed mass, which increases on reducing the coupling and enlarging $\Lambda_{\mathrm UV}$~\cite{Bedolla:2015mpa}. Furthermore, we see that, as long as we reduce $\Lambda_{\mathrm UV}$, the mass gap between the pseudoscalar and vector channels decreases slightly; it is therefore expected that this ordering reverses at some point, as can be seen in this figure. A similar behaviour also appears in Ref.~\cite{Qin:2011xq}: when the infrared length scale of the model therein increases, the ordering is reversed and a correct ordering $m_{\pi}^*>m_{\rho}^*$ is presented.

\begin{table}[h]
\begin{center}
\begin{tabular}{lllll}
\hline \hline
    & \multicolumn{4}{c}{masses [GeV]}  \\
\hline
& $m_{\pi(1S)}$ & $m_{rho(1S)}$ & $m_{\sigma(1P)}$ & $m_{a_1(1P)}$
\\
\hline
Experiment~\cite{Olive:2016xmw} & 0.140 & 0.780 & 1.0-1.2 & 1.230  \\
CI~\cite{Bedolla:2015mpa} & 0.140 & 0.93 & 1.29 & 1.38 \\
\hline
    & \multicolumn{4}{c}{first radial excitation}  \\
\hline
Experiment~\cite{Olive:2016xmw} & 1.3 & 1.47 & -- & 1.65  \\
CI~\cite{Roberts:2011wy,Chen:2012qr} & 1.33 & 1.29 & 1.43 & 1.47 \\
SQ~\cite{Qin:2011xq} & 1.283 & 1.260 & 1.489 & --- \\
\hline \hline
\end{tabular}
\caption{\label{tab:muu_all_opt} Ground-state and first radial excitation light-sector mass spectrum.
  The CI results were obtained with the best-fit parameter set: $m_{g}= 0.8\,\GeV$,
  $\alpha_{IR}= 0.93\pi$, $\Lambda_{\text{IR}}= 0.24\,\GeV$, and $\Lambda_{\text{UV}}= 0.905\,\GeV$.
  The current-quark mass is $m_{u}= 0.007\,\GeV$, and the dynamically generated constituent-like
  mass is $M_{u}= 0.37\,\GeV$.}
\end{center}
\end{table}
%
%
%

\begin{table}[h]
\begin{center}
\begin{tabular}{lllll}
\hline \hline
    & \multicolumn{4}{c}{masses [GeV]}  \\
\hline
& $m_{\eta_{c}(1S)}$ & $m_{J/\Psi(1S)}$ & $m_{\chi_{c_{0}}(1P)}$ & $m_{\chi_{c_{1}}(1P)}$
\\
\hline
Experiment~\cite{Olive:2016xmw} & 2.983 & 3.096 & 3.414 & 3.510  \\
CI~\cite{Bedolla:2015mpa} & 2.950 & 3.129 & 3.407 & 3.433 \\
\hline
    & \multicolumn{4}{c}{first radial excitation}  \\
\hline
Experiment~\cite{Olive:2016xmw} & 3.639 & 3.686 & --- & ---  \\
CI & 3.135 & 3.123 & 3.375 & 3.401 \\
CI* & 3.600 & 3.565 & 3.780 & 3.803 \\
JM~\cite{Jain:1993qh} & 3.470 & 3.700 & --- & --- \\
RB1 & 3.402\cite{Rojas:2014aka} & 3.393\cite{Mojica:2017tvh} & --- & --- \\
RB2 & 3.784\cite{Rojas:2014aka} & 3.507\cite{Mojica:2017tvh} & --- & ---  \\
FKW~\cite{Fischer:2014cfa} & 3.683 & 3.676 & 3.833 & 3.673 \\
HGKL1~\cite{Hilger:2017jti} & 3.508 & 3.553 & --- & 3.523 \\ 
HGKL2~\cite{Hilger:2017jti} & 3.384 & 3.438 & 3.575 & 3.420 \\ 
\hline \hline
\end{tabular}
\caption{\label{tab:mcc_all_opt} Ground-state and first radial excitation charmonium mass spectrum.
  The CI results were obtained with the best-fit parameter set: $m_{g}= 0.8\,\GeV$,
  $\alpha_{\mathrm IR}= 0.93\pi/17$, $\Lambda_{\text{IR}}= 0.24\,\GeV$, and $\Lambda_{\text{UV}}= 2.4\,\GeV$.
  The current-quark mass is $m_{c}= 1.09\,\GeV$, and the dynamically generated constituent-like
  mass is $M_{c}= 1.481\,\GeV$. *Results obtained with parameters adjusted to give the mass of $\eta_c (2S)$: $\alpha_{\mathrm IR}= 0.93\pi/44.09$, $\Lambda_{\text{UV}}= 3.761\,\GeV$, $M_{c}=1.690\,\GeV$.}
\end{center}
\end{table}
%
\begin{table}[h]
\begin{center}
\begin{tabular}{lllll}
\hline \hline
    & \multicolumn{4}{c}{masses [GeV]}  \\
\hline
& $m_{\eta_{b}(1S)}$ & $m_{\Upsilon(1S)}$ & $m_{\chi_{b_{0}}(1P)}$ & $m_{\chi_{b_{1}}(1P)}$
\\
\hline
Experiment~\cite{Olive:2016xmw} & 9.32 (9.4) & 9.46 & 9.860 & 9.892 \\
CI & 9.407 & 9.547 & 9.671 & 9.680 \\ 
\hline
    & \multicolumn{4}{c}{first radial excitation}  \\
\hline
Experiment~\cite{Olive:2016xmw} & 9.999 & 10.023 & 10.232 & 10.255 \\
CI& 9.482 & 9.480 & 9.594 & 9.603 \\ 
CI* & 9.992 & 9.982 & 10.081 & 10.090 \\
JM~\cite{Jain:1993qh} & 9.770 & 9.980	 & --- & --- \\
RB1~\cite{Mojica:2017tvh} & --- & 9.945 & --- & --- \\
RB2~\cite{Mojica:2017tvh} & --- & 9.848 & --- & ---  \\
FKW~\cite{Fischer:2014cfa} & 9.987 & 10.089 & 10.254 & 10.120 \\ 
HGKL1~\cite{Hilger:2017jti} & 9.820 & 9.838 & 10.004 & 9.790 \\ 
HGKL2~\cite{Hilger:2017jti} & 9.728 & 9.754 & 9.917 & 9.761 \\ 
\hline \hline
\end{tabular}
\caption{\label{tab:mbb_all_opt} Ground-state bottomonium mass spectrum.
  The CI results were obtained with the best-fit parameter set: $m_{g}= 0.8\,\GeV$,
  $\alpha_{IR}= 0.93\pi/125$, $\Lambda_{\text{IR}}= 0.24\,\GeV$, and $\Lambda_{\text{UV}}= 6.4\,\GeV$.
  The current-quark mass is $m_{b}= 3.8\,\GeV$, and the dynamically generated constituent-like
  mass is $M_{b}= 4.710\,\GeV$. *Results obtained with parameters adjusted to give the mass of $\eta_b (2S)$: $\alpha_{IR}= 0.93\pi/458.567$, $\Lambda_{\text{UV}}= 10.17\,\GeV$, $M_{b}= 4.959\,\GeV$.}
\end{center}
\end{table}
%

\subsection{\label{sec:Charmoniaradialexcitations} Charmonium first radial excitation}

\begin{figure}[ht]
\includegraphics[width=0.45\textwidth]{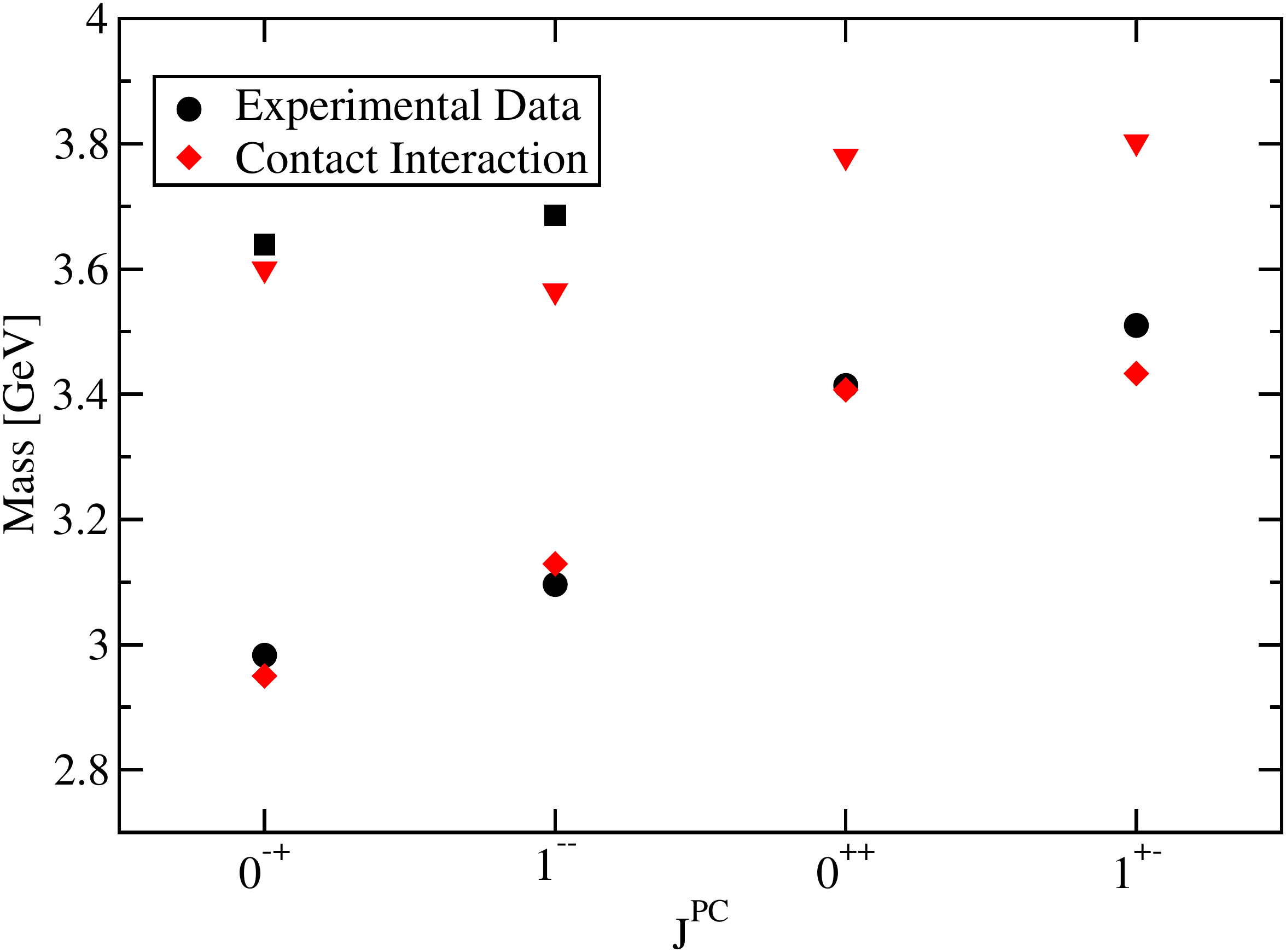}
\caption{\label{fig:charmoniaspectra} Contact interaction results for the $c\bar c$ mass spectrum; see \tab{tab:mcc_all_opt}. Experimental data are taken from Ref.~\cite{Olive:2016xmw}.}
\end{figure}
%
In \tab{tab:mcc_all_opt}, we present new results for charmonia first radial excitations with new parameters. We propose fixing $\Lambda_{\mathrm UV}$ in order to obtain $m_{\eta_c (2S)}=3.6\,\mathrm{GeV}$, while the remaining channels are predicted values. We found that $m_{\chi_{c0}(2P)}>m_{\chi_{c1} (2P)}$ , which is in contrast with other model predictions. Even though there are not experimental results for $m_{\chi_{c0}(2P)}$ and $m_{\chi_{c1}(2P)}$, we expect a good comparison when data become available. In \fig{fig:charmoniaspectra}, we present a pictographical scheme of the spectra and compare this with experimental data when they are available.

Before continuing to the analysis of bottomonia results, it is important to consider the $X(3930)$~\cite{Abe:2004zs} and $X(3915)$~\cite{Uehara:2009tx} states, both with $J^{PC}=0^{++}$~\cite{Lees:2012xs}, which suggests that one of them could be the $\chi_{{c0}(2P)}$ quark model state~\cite{Zhou:2015uva}, though there are still some details to explain. Nevertheless, our constant interaction model suggests a mass splitting $m_{\chi_{{c0}(2P)}}-m_{\chi_{{c1}(2P)}}=13$ MeV, while other models predict $m_{\chi_{{c1}(2P)}}-m_{\chi_{{c0}(2P)}}=>100$ MeV. In addition, it is necessary to stress that there is a switched ordering between the scalar and the axial-vector channel predicted by means of the CI model and fully covariant models. Moreover, in considering bottomonium experimental results, we consider our results to be better approximation and an acceptable hint for the $m_{\chi_{{c0}(2P)}}-m_{\chi_{{c1}(2P)}}$ mass splitting.

\subsection{\label{sec:Bottomoniaradialexcitations} Bottomonium first radial excitation}

\begin{figure}[ht]
\includegraphics[width=0.45\textwidth]{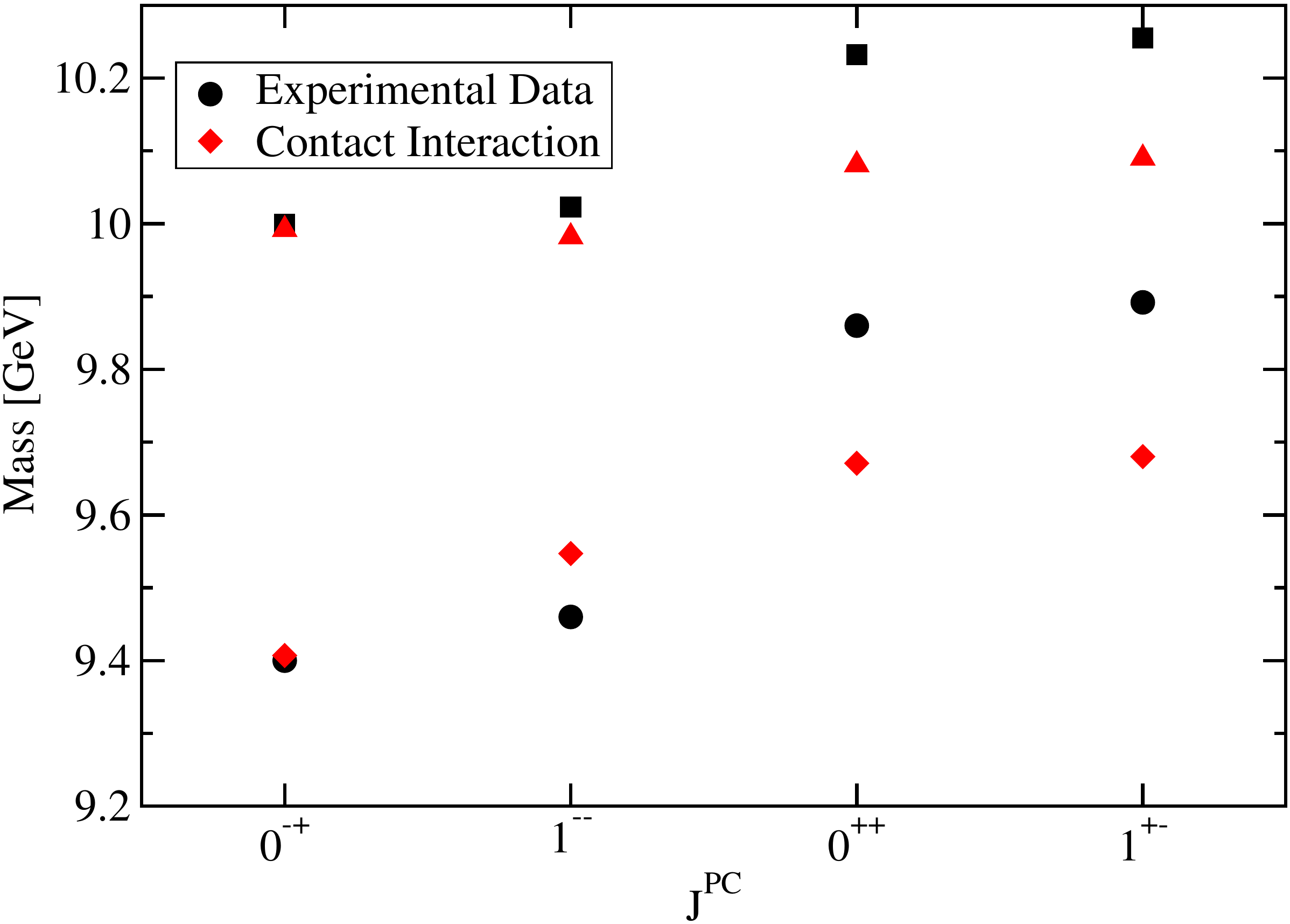}
\caption{\label{fig:bottomoniaspectra} Contact interaction results for the $b\bar b$ mass spectrum; see \tab{tab:mbb_all_opt}. Experimental data are taken from Ref.~\cite{Olive:2016xmw}.}
\end{figure}
Conclusions for bottomonia are a repetition of the results for charmonia. \tab{tab:mbb_all_opt} shows our previous results for ground-state mass spectra in four different channels. These values are in excellent agreement with experimental data~\cite{Olive:2016xmw} and those of other model calculations. However, when calculating the first radial excitation, we again have a drawback: the masses of excited states are lower than their ground-state counterparts. \fig{fig:bottomoniaspectra} shows the bottomonium spectrum with the value of $\Lambda_{\mathrm UV}=10.17\GeV$ fixed to obtain $m_{\Upsilon(2S)}=9.992\GeV$; the rest of the predicted results are presented in \tab{tab:mbb_all_opt}. In this case, the experimental mass splitting $m_{\chi_{{b0}(2P)}}-m_{\chi_{{b1}(2P)}}=23$ MeV is in agreement with our value ($9$ MeV), whereas other models predict $m_{\chi_{{b1}(2P)}}-m_{\chi_{{b0}(2P)}}>150$ MeV. Additionally, the CI model predicts the correct mass ordering between these particles, in contrast to other models, which predict a reversed ordering.

Finally, from \fig{fig:bottomoniaspectra} and \tab{tab:mbb_all_opt}, we conclude that, when comparing the mass splitting $m_{\chi_{{b0}(2P)}}-m_{\chi_{{b0}(1P)}}$ and $m_{\chi_{{b0}(2P)}}-m_{\chi_{{b0}(1P)}}$  between the experimental and those of our model, they match perfectly. These results suggest that the CI predicted values in Sec.~\ref{sec:Charmoniaradialexcitations} are close data to the experimental ones.

\subsection{\label{sec:DCradialexcitations} Decay constants}

\begin{table}[h]
\begin{center}
\begin{tabular}{lllll}
\hline \hline
    & \multicolumn{3}{c}{amplitudes}  \\
\hline \hline
& $f_{\eta_{c}(1S)}$ & $f_{J/\Psi(1S)}$ & $f_{\chi_{c_{0}}(1P)}$ & $f_{\chi_{c_{1}}(1P)}$
\\
$E_H$ & 1.544 & 0.477 & 0.033 & 0.084 \\
$F_H$ & 0.281 & ---  & --- & ---\\
\hline     & \multicolumn{4}{c}{first radial excitation}  \\
\hline
\hline
& $f_{\eta_{c}(2S)}$ & $f_{\Psi(2S)}$ & $f_{\chi_{c_{0}}(2P)}$ & $f_{\chi_{c_{1}}(2P)}$
\\
$E_H$ & 0.380 & 0.212 & 0.021 & 0.015 \\
$F_H$ & 0.123 & ---  & --- & ---\\
\hline \hline
\end{tabular}
\caption{\label{tab:charmAM} Ground-state and first radially excited canonically normalized amplitudes of charmonia. Ground-state amplitudes were obtained  with the first set of parameters found in \tab{tab:mcc_all_opt}, while radially excited states were obtained with the second set.}
\end{center}
\end{table}

\begin{table}[h]
\begin{center}
\begin{tabular}{lllll}
\hline \hline
    & \multicolumn{3}{c}{decay constants [GeV]}  \\
\hline \hline
 & $f_{\eta_{c}(1S)}$ & $f_{J/\Psi (1S)}$
\\
Experiment~\cite{Olive:2016xmw} & 0.238 & 0.294 \\
CI~\cite{Bedolla:2015mpa} & 0.255 & 0.206  \\
\hline     & \multicolumn{4}{c}{first radial excitation}  \\
\hline
\hline
 & $f_{\eta_{c}(2S)}$ & $f_{\Psi (2S)}$
\\
Experiment~\cite{Olive:2016xmw} & --- & 0.208 \\
CI[This work] & 0.296 & 0.205 \\
RB1~\cite{Rojas:2014aka} & 0.063 & 0.147  \\
RB2~\cite{Rojas:2014aka} & 0.062 & 0.162  \\
HGKL1~\cite{Hilger:2017jti} & 0.103 & 0.096  \\ 
HGKL2~\cite{Hilger:2017jti} & 0.089 & 0.121  \\ 
\hline \hline
\end{tabular}
\caption{\label{tab:charmDC} Decay constants of $\eta_{c}(1S,2S)$, $J/\Psi(1S)$ and  $\Psi(2S)$. Note that the numerical values contain a division by $\sqrt{2}$ for a consistent comparison between different computations. CI results were obtained with the parameter set used to produce \tab{tab:mcc_all_opt}.}
\end{center}
\end{table}
%
\begin{table}[h]
\begin{center}
\begin{tabular}{lllll}
\hline \hline
    & \multicolumn{3}{c}{amplitudes}  \\
\hline \hline
& $f_{\eta_{b}(1S)}$ & $f_{\Upsilon(1S)}$ & $f_{\chi_{b_{0}}(1P)}$ & $f_{\chi_{b_{1}}(1P)}$
\\
$E_H$ & 1.311 & 0.211 & 0.012 & 0.030 \\
$F_H$ & 0.265 & ---  & --- & ---\\
\hline     & \multicolumn{4}{c}{first radial excitation}  \\
\hline
\hline
& $f_{\eta_{b}(2S)}$ & $f_{\Upsilon(2S)}$ & $f_{\chi_{b_{0}}(2P)}$ & $f_{\chi_{b_{1}}(2P)}$
\\
$E_H$ & 0.302 & 0.153 & 0.012 & 0.008 \\
$F_H$ & 0.078 & ---  & --- & ---\\
\hline \hline
\end{tabular}
\caption{\label{tab:bottomAM} Ground-state and first radially excited canonically normalized amplitudes of bottomonia. Ground-state amplitudes were obtained  with the first set of parameters found in \tab{tab:mbb_all_opt}, while radially excited states were obtained with the second set.}
\end{center}
\end{table}
	
\begin{table}[h]
\begin{center}
\begin{tabular}{lllll}
\hline \hline
    & \multicolumn{3}{c}{decay constants [GeV]}  \\
\hline \hline
 & $f_{\eta_{b}(1S)}$ & $f_{\Upsilon (1S)}$
\\
Experiment~\cite{Olive:2016xmw} & --- & 0.506 \\
CI  & 0.553 & 0.219 \\
\hline
\hline
 & $f_{\eta_{b}(2S)}$ & $f_{\Upsilon (2S)}$
\\
Experiment~\cite{Olive:2016xmw} & --- & 0.341 \\
CI[This work] & 0.163 & 0.157 \\
Lattice &-- & 0.340~\cite{Colquhoun:2014ica} \\
HGKL1~\cite{Hilger:2017jti} & 0.186 & 0.200  \\ 
HGKL2~\cite{Hilger:2017jti} & 0.207 & 0.230  \\ 
\hline \hline
\end{tabular}
\caption{\label{tab:bottomDC} Decay constants of $\eta_{b}(1S,2S)$ and $\Upsilon(1S,2S)$. Note that the numerical values contain a division by $\sqrt{2}$ for a consistent comparison between different computations. CI results were obtained with the parameter set used to produce \tab{tab:mbb_all_opt}.}
\end{center}
\end{table}

Charmonia BSAs are displayed in \tab{tab:charmAM}. First radial excitation obtained by considering the set of parameters presented in \tab{tab:mcc_all_opt}, whereas $\eta_c (2S)$ and $\Psi (2S)$ decay constants are reported in \tab{tab:charmDC}. Similarly, bottomonia BSAs are given in \tab{tab:bottomAM}, and $\eta_b (2S)$ and $\Upsilon(2S)$ decay constants are shown in \tab{tab:bottomDC}. It is important to notice that, the canonically normalized amplitude associated with the excited states BSE, are smaller than their corresponding of the ground-states. Otherwise, our results are in reasonably good agreement with experimental results when they are available. 

However, the value of the $\eta_c (2S)$ decay constant predicted by our model differs too much from those of other model predictions; these latter present values from 0.063 to 0.103, while our predicted value is 0.296, which is even greater than the $\eta_c (1S)$ decay constant. We attribute this enormous difference to the fact that the decay constant is sensitive to changes in the model coupling and ultraviolet cut-off $\Lambda_{\rm UV}$. It is also important to consider, that this, time we only fitted the parameters to the value of $m_{\eta_c (2S)}$. On the other hand, the $\Psi (2S)$ decay constant is in agreement with the experimental value. The bottomonia scene presents a better picture; the predicted values for $\eta_b (2S)$ and $\Upsilon(2S)$ are within 10\% concordance with other model predictions, but all of them halve the experimental value. It is  expected that future results beyond RL approximation will improve the results of decay constants.

\section{\label{sec:conclusions} Conclusions}

\noindent We compute the quark model first radial excitation state spin-0 and spin-1 heavy quarkonia masses, and the decay constants by using a rainbow-ladder approximation of the simultaneous set of SDE and BSE. By means of a
CI model of QCD previously developed to the study of ground-state mesons~\cite{GutierrezGuerrero:2010md,Roberts:2010rn,Roberts:2011cf,Roberts:2011wy,Chen:2012qr,Bedolla:2015mpa,Raya:2017ggu}. 

In a fully covariant approach, a radial excitation is a zero in the relative-momentum dependence of the leading Tchebychev momentum of its dominant Dirac structure. This property precludes the possibility that the interaction between quarks is momentum-independent, as an independent relative-momentum bound-state amplitude cannot exhibit a single zero. However, we have inserted a zero by hand into the analytical expressions of the BSE. In this way, the coupling is reduced and the mass of the bound state increases. This idea was first implemented in calculating the spectra of light-mesons first radial excitation. Although, the results provided a good comparison with experimental data, the mass splitting between their ground state and first radially excited state was underestimated by $0.2$ GeV~\cite{Chen:2012qr}. On applying this same idea to heavy quarkonia, we obtained a first radial excited state that was too low for a correct comparison with experimental data, and changing the new parameter does not improve this picture.

In our previous studies of heavy-quarkonia in a contact interaction~\cite{Bedolla:2015mpa,Raya:2017ggu}, we showed that the extension of the CI model to heavy quarkonia requires a reduction in the model interaction strength, mimicking the high momentum tail of the quark mass function and BSAs~\cite{Bhagwat:2004hn,Maris:1997tm,Maris:1999nt} as a consequence
of the asymptotic freedom of QCD. This reduction in the interaction strength is appropriately
compensated by increasing the ultraviolet cut-off; from these parameters, a dimensionless
interaction strength can be defined. With this in mind, we have unified a contact interaction model to study light and heavy mesons; we have also recognized that the coupling can be fitted reasonably well with a inverse logarithmic curve, with an appearance evocative of the QCD running coupling.

In order to obtain a correct mass of the first radially excited state, we suggest that heavy excited mesons need different parameters from those of their ground-state counterparts. This time, instead of use a best fit to obtain the correct values of the pseudoscalar channel, we used the inverse logarithmic curve to obtain a value of the coupling CI model, $\alpha_{\mathrm IR}$, in terms of the ultraviolet cut-off $\Lambda_{\mathrm UV}$. This time we find that the masses of the first radially excited-state heavy quarkonium are in good agreement with experimental data~\cite{Olive:2016xmw}. However, we find an incorrect ordering between the pseudoscalar and vector channel, which can be mainly attributed to a defect of the RL approximation.

One important feature of our model is that predicts a mass splitting between $m_{\chi_{{b0}(2P)}}-m_{\chi_{{b0}(1P)}}$ and $m_{\chi_{{b1}(2P)}}-m_{\chi_{{b1}(1P)}}$ that is consistent with experiment, though each value is underestimated  by $\approx 200$MeV in relation to experiment. Consequently, if we think the mass splitting values $m_{\chi_{{c0}(2P)}}-m_{\chi_{{c0}(1P)}}$ and $m_{\chi_{{c1}(2P)}}-m_{\chi_{{c1}(1P)}}$ are correct, and as charmonia ground-state mass values are close to the experimental ones, we can state that the values of $m_{\chi_{{c0}(2P)}}$ and $m_{\chi_{{c1}(2P)}}$ might be close to experimental results once these states are detected.

Finally, the decay constants that we have obtained are rather inconsistent in the charmonia sector, our predictions predictions differ too much from those of other model predictions for $\eta_c (2S)$. Nonetheless, we have an excellent agreement with experiment in the value for $\Psi (2S)$. On the other hand, the predicted results for bottomonia are pretty similar for both $\eta_b (2S)$ and $\Upsilon (2S)$, but there is a mismatch of almost 50\% in the predicted value for $\Upsilon (2S)$. We expect that future studies beyond RL approximation will improve these values.

This work is part of the series of studies on heavy-quarkonia in a contact interaction~\cite{Bedolla:2015mpa,Bedolla:2016yxq,Serna:2017nlr,Raya:2017ggu}, in which we move towards a comprehensive study of heavy mesons and QCD by using this model. Further steps will involve flavored mesons and baryons, as well as exotic states. Our goal is to provide a unified phenomenological description of light and heavy hadrons within the CI model.

\section{Acknowledgments}

The authors acknowledge financial support from CONACyT, M\'exico (postdoctoral
fellowship for M.A.~Bedolla), the INFN Sezione di Genova and the Instituto de Fisica y Matematicas.

\appendix
\section{\label{App:ExcitedRegularization} Excited Kernel Extra Terms}

This appendix provides the explicit regularized expressions of the excited kernel extra terms given in Eqs.~(\ref{eqn:excited_reg},~\ref{eqn:excited_reg2}). 

We solve \eqn{eqn:excited_reg} after the substitution $s'\to s+\mathfrak{M}^{2}$, and obtain
\begin{IEEEeqnarray}{rCl}
{\cal D}_{01}(\mathfrak{M}^{2}) & = & \mathfrak{M}^{4}\,\Gamma\left(0,\tau_{\text{UV}}^2 \mathfrak{M}^{2},\tau_{\text{IR}}^2 \mathfrak{M}^{2}\right)\nonumber\\
& & -2 \mathfrak{M}^{2}\left(\frac{e^{-\mathfrak{M}^{2}\tau_{\text{IR}}^2}}{\tau_{\text{IR}}^2}- \frac{e^{-\mathfrak{M}^{2}\tau_{\text{UV}}^2}}{\tau_{\text{UV}}^2}\right)
\nonumber\\
&&+ \frac{e^{-\mathfrak{M}^{2}\tau_{\text{UV}}^2}}{\tau_{\text{UV}}^4}-\frac{e^{-\mathfrak{M}^{2}\tau_{\text{IR}}^4}}{\tau_{\text{IR}}^4}\,.\label{eqn:excited_reg_exp}
\end{IEEEeqnarray}
The rest of the expressions are obtained through differentiation
\begin{IEEEeqnarray}{rCl}
{\cal D}_{02}(\mathfrak{M}^{2}) & = &-\frac{d}{d \mathfrak{M}^2} {\cal D}_{01}(\mathfrak{M}^{2})\nonumber\\
& =& +2 \mathfrak{M}^{2} \left(\frac{e^{-\mathfrak{M}^{2}\tau_{\text{UV}}^2}}{\tau_{\text{UV}}^2}- \frac{e^{-\mathfrak{M}^{2}\tau_{\text{IR}}^2}}{\tau_{\text{IR}^2}}\right.
\nonumber\\
&& + \mathfrak{M}^{2}\Gamma\left(0, \mathfrak{M}^{2}\tau_{\text{UV}}^2,\mathfrak{M}^{2}\tau_{\text{IR}}^2\right)\Bigg)\,.\label{eqn:excited_reg_exp02}\\
{\cal D}_{03}(\mathfrak{M}^{2}) & = &-\frac{d}{d \mathfrak{M}^2} {\cal D}_{02}(\mathfrak{M}^{2})\nonumber\\
&=& 2 \Gamma\left(0, \mathfrak{M}^{2}\tau_{\text{UV}}^2,\mathfrak{M}^{2}\tau_{\text{IR}}^2\right)\,.\label{eqn:excited_reg_exp03}
\end{IEEEeqnarray}
\eqn{eqn:excited_reg2} is solved analogously
\begin{IEEEeqnarray}{rCl}
{\cal E}_{01}(\mathfrak{M}^{2}) & = & -\mathfrak{M}^{6}\,\Gamma\left(0,\tau_{\text{UV}}^2 \mathfrak{M}^{2},\tau_{\text{IR}}^2 \mathfrak{M}^{2}\right)\nonumber\\
& & +3 \mathfrak{M}^{4}\left(\frac{\Gamma\left(1,\tau_{\text{UV}}^2 \mathfrak{M}^{2}\right)}{\tau_{\text{UV}}^2}- \frac{\Gamma\left(1,\tau_{\text{IR}}^2 \mathfrak{M}^{2}\right)}{\tau_{\text{IR}}^2}\right)
\nonumber\\
& & -3 \mathfrak{M}^{2}\left(\frac{\Gamma\left(2,\tau_{\text{UV}}^2 \mathfrak{M}^{2}\right)}{\tau_{\text{UV}}^4}- \frac{\Gamma\left(2,\tau_{\text{IR}}^2 \mathfrak{M}^{2}\right)}{\tau_{\text{IR}}^4}\right)
\nonumber\\
&&+ \frac{\Gamma\left(3,\tau_{\text{UV}}^2 \mathfrak{M}^{2}\right)}{\tau_{\text{UV}}^6}- \frac{\Gamma\left(3,\tau_{\text{IR}}^2 \mathfrak{M}^{2}\right)}{\tau_{\text{IR}}^6}\,.\label{eqn:excited_reg2_exp}
\end{IEEEeqnarray}
Analogously, the remaining terms are obtained through differentiation
\begin{IEEEeqnarray}{rCl}
{\cal E}_{02}(\mathfrak{M}^{2}) & = &-\frac{d}{d \mathfrak{M}^2} {\cal E}_{01}(\mathfrak{M}^{2})\nonumber\\
& =&  6 \,\mathfrak{M}^{2} \left(\frac{e^{-\mathfrak{M}^{2}\tau_{\text{UV}}^2}}{\tau_{\text{UV}}^2}- \frac{e^{-\mathfrak{M}^{2}\tau_{\text{IR}}^2}}{\tau_{\text{IR}^2}}\right)
\nonumber\\
&& -3 \,\mathfrak{M}^{4}\Gamma\left(0, \mathfrak{M}^{2}\tau_{\text{UV}}^2,\mathfrak{M}^{2}\tau_{\text{IR}}^2\right)\nonumber\\
&& -3\,\left(\frac{\Gamma\left(2,\tau_{\text{UV}}^2 \mathfrak{M}^{2}\right)}{\tau_{\text{UV}}^4}- \frac{\Gamma\left(2,\tau_{\text{IR}}^2 \mathfrak{M}^{2}\right)}{\tau_{\text{IR}}^4} \right)\,.\label{eqn:excited_reg2_exp02}\\
{\cal E}_{03}(\mathfrak{M}^{2}) & = &-\frac{d}{d \mathfrak{M}^2} {\cal E}_{02}(\mathfrak{M}^{2})\nonumber\\
& =&  6 \,\left(\frac{e^{-\mathfrak{M}^{2}\tau_{\text{UV}}^2}}{\tau_{\text{UV}}^2}- \frac{e^{-\mathfrak{M}^{2}\tau_{\text{IR}}^2}}{\tau_{\text{IR}^2}}\right)
\nonumber\\
&=& -6\, \Gamma\left(0, \mathfrak{M}^{2}\tau_{\text{UV}}^2,\mathfrak{M}^{2}\tau_{\text{IR}}^2\right)\,.\label{eqn:excited_reg2_exp03}
\end{IEEEeqnarray}

\end{document}